\documentclass[prl,aps,twocolumn]{revtex4}
\usepackage{epsfig,amsmath,amsfonts}
\newcommand{\nc}{\newcommand}
\nc{\rnc}{\renewcommand}
\nc{\ket}[1]{| #1 \rangle}
\nc{\bra}[1]{\langle #1 |}
\nc{\proj}[1]{\ket{#1}\bra{#1}}
\nc{\braket}[2]{\langle #1| #2 \rangle}
\nc{\hilb}{\mathcal{H}}
\nc{\inprod}[2]{\braket{#1}{#2}}

\rnc{\vec}[1]{\boldsymbol{#1}}
\begin{document}

\title{Generalization of geometric phase to completely positive maps}

\author{Marie Ericsson}
\author{Erik Sj\"{o}qvist} 
\address{Department of Quantum Chemistry, 
Uppsala University, Box 518, Se-751 20, Sweden}
\author{Johan Br\"annlund}
\address{Stockholm University, SCFAB, Fysikum, 
Se-106 91 Stockholm, Sweden}
\author{Daniel K.L. Oi}
\address{Centre for Quantum Computation, Clarendon Laboratory, University 
of Oxford, Parks Road, Oxford OX1 3PU, UK}
\author{Arun K. Pati}
\address{Institute of Physics, Bhubaneswar-751005 Orissa, India}

\begin{abstract}
  We generalize the notion of relative phase to completely positive maps with
  known unitary representation, based on interferometry.  Parallel transport
  conditions that define the geometric phase for such maps are introduced. The
  interference effect is embodied in a set of interference patterns defined by
  flipping the environment state in one of the two paths. We show for the
  qubit that this structure gives rise to interesting additional 
  information about the geometry of the evolution defined by the CP map.
\end{abstract} 
\maketitle 

Berry's \cite{berry84} discovery of a geometric phase accompanying cyclic
adiabatic evolution has triggered an immense interest in holonomy effects in
quantum mechanics and has led to many generalizations. The restriction of
adiabaticity was removed by Aharonov and Anandan \cite{aharonov87}, who
pointed out that the geometric phase is due to the curvature of the projective
Hilbert space. It was extended to noncyclic evolution by Samuel and Bhandari
\cite{samuel88} (see also Refs. \cite{aitchison92}), based on Pancharatnam's
\cite{pancharatnam56} work on interference of classical light in distinct
states of polarization. Another development of geometric phase was initiated
by Uhlmann \cite{uhlmann86} who introduced this notion to mixed quantal
states. More recently another mixed state geometric phase in the particular
case of unitary evolution was discovered in the context of interferometry
\cite{sjoqvist00}.
 
The geometric phase has shown to be useful in the context of quantum computing
as a tool to achieve fault-tolerance \cite{jones00}. For practical
implementations of geometric quantum computing, it is important to understand
the behavior of the geometric phase in the presence of decoherence.  For this,
we generalize in this Letter the idea in \cite{sjoqvist00} to completely
positive (CP) maps, i.e. we define the relative (Pancharatnam) phase and
introduce a notion of parallel transport with concomitant geometric phase for
such maps. These generalized concepts reduces to that of \cite{sjoqvist00} in
the case of unitary evolutions.
 
Let us first consider a Mach-Zehnder interferometer with a variable 
relative $U(1)$ phase $\chi$ in one of the interferometer beams (the 
reference beam) and assume that the interfering particles carry an 
additional internal degree of freedom, such as spin or polarization, 
in a pure state $|k\rangle$. The other beam (the target beam) is  
exposed to the unitary operator $U_{i}$, yielding the output 
interference pattern $I \propto 1+\nu\cos(\chi-\alpha)$, which 
is completely determined by the complex quantity  
\begin{equation} 
\nu e^{i\alpha}=\langle k| U_{i}|k\rangle. 
\label{eq:purequantity} 
\end{equation} 
Thus, by varying $\chi$, the relative phase $\alpha$ and visibility  
$\nu$ can be distinguished experimentally. We note that $\alpha$ is  
a shift in the maximum of the interference pattern, a fact that  
motivated Pancharatnam \cite{pancharatnam56} to define it as  
the relative phase between the internal states $\ket{k}$ and  
$U_{i}\ket{k}$ of the two beams.  
 
Pancharatnam's analysis was generalized in \cite{sjoqvist00} 
to mixed states undergoing unitary evolution as follows. Assume 
that the incoming particle is in a mixed internal state 
$\rho=\sum_{k=1}^{N}w_{k}|k\rangle\langle k|$, where $N$ is 
the dimension of the internal Hilbert space. Each pure 
component $|k\rangle$ of this mixture contributes an  
interference profile given by $\langle k|U_{i}|k\rangle =  
\nu_{k} e^{i\alpha_{k}}$ weighted by the its  
probability $w_{k}$ yielding $I=\sum_{k} w_{k} I_{k} \propto 1 +  
\sum_{k}w_{k}\nu_{k} \cos[\chi-\alpha_{k}]$. Noting that 
${\text{Tr}}(U_{i}\rho) = \sum w_{k}\langle k|U_{i}|k \rangle$, 
this can also be written as $I \propto 1 + |{\text{Tr}} 
(U_{i}\rho)| \cos[\chi - \arg {\text{Tr}}(U_{i}\rho)]$.  
The key result is that the interference fringes,
produced by varying the phase $\chi$, is shifted by  
$\alpha=\arg{\text{Tr}}(U_{i} \rho)$ and that this shift reduces  
to Pancharatnam's original prescription for pure states. These  
two facts are the central properties for $\alpha$ being a  
natural generalization of Pancharatnam's relative phase to  
mixed states undergoing unitary evolution. Furthermore, it  
is clear that the quantity that extends that of Eq.~(\ref{eq:purequantity}) 
to mixed states is
\begin{equation} 
\nu e^{i\alpha}={\text{Tr}} (U_{i}\rho), 
\label{eq:int2} 
\end{equation} 
with visibility $\nu = |{\text{Tr}} (U_{i}\rho) |$.  
 
Nonunitary evolution of a quantal state may be conveniently  
modeled by appending an environment in a pure state that we 
designate $|0_e\rangle$, i.e.   
\begin{equation}
\varrho=\rho \otimes |0_{e}\rangle 
\langle 0_{e}|, 
\label{eq:rho}  
\end{equation} 
and letting the combined state evolve unitarily as   
$\varrho \rightarrow \varrho' = U_{ie}\varrho U^{\dagger}_{ie}$ 
with given $U_{ie}$. The evolved density matrix  
of the internal part is obtained by tracing over this   
environment yielding    
\begin{equation} 
\rho'={\text{Tr}}_{e} \varrho'=\sum_{\mu} 
m_{\mu}\rho m_{\mu}^{\dagger}, 
\label{eq:kraus}
\end{equation} 
where the Kraus operators are $m_{\mu} = \langle\mu_{e} |U_{ie}|0_{e} \rangle$
\cite{kraus83} in terms of an orthonormal basis $\{|\mu_{e}\rangle\}$, $\mu =
0, \ldots, K-1 \geq N$, of the $K-$dimensional Hilbert space of the
environment. This map $\Lambda$ is completely positive (CP), i.e. it takes
density operators into density operators, and all trivial extensions $I\otimes\Lambda$
likewise. Conversely, any CP map has a Kraus representation of the form
Eq.~(\ref{eq:kraus}).
 
Using Eq.~(\ref{eq:int2}), the interference pattern  
for the incoming state in Eq.~(\ref{eq:rho}), evolved with  
$U_{ie}$ in the target beam, is described by  
\begin{eqnarray} 
\nu_{0} e^{i\alpha_{0}} & = &  
{\text{Tr}}_{ie} \big[ U_{ie} \varrho \big] =  
{\text{Tr}}_{i}\big[m_{0}\rho\big], 
\label{eq:cpint} 
\end{eqnarray} 
where we have used $\langle 0_{e} | \mu_{e} 
\rangle = \delta_{0\mu}$. The quantity $\alpha_{0}$ is a natural  
definition of relative phase as it shifts the maximum of the  
interference pattern and reduces to the phase defined in  
\cite{sjoqvist00} for unitarily evolving mixed states.  

\begin{figure}
\epsfig{figure=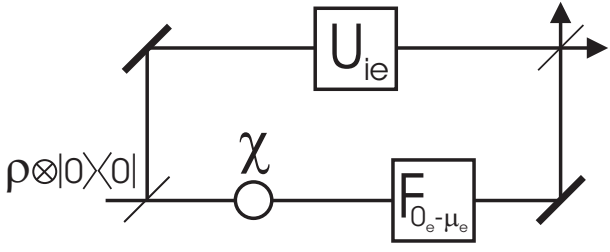,width=0.45\textwidth}
\caption{Interferometer for determining complete interference information.}
\label{fig:CPint}
\end{figure}

\begin{figure}
\epsfig{figure=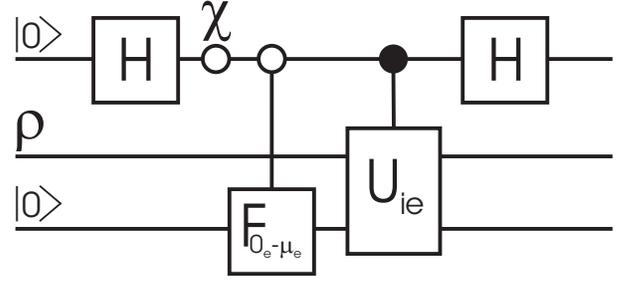,width=0.45\textwidth}
\caption{Quantum network for interferometry of a state undergoing a CP map.}
\label{fig:CPNetwork}
\end{figure}

Since phase information has leaked from the system part, the interference
information contained in Eq. (\ref{eq:cpint}) is only partial. The remaining
part may be uncovered by flipping the state of the environment associated with
the reference beam to an orthogonal state $|\mu_{e} \neq 0_e \rangle$
(Fig.~\ref{fig:CPint}) \footnote{This assumes we have full control over the
environment and that it is, e.g., an extra degree of internal freedom.}.
This transformation may be represented by the operator
\begin{eqnarray} 
U=\left( \begin{array}{rr} 0 & 0 \\ 0 & 1  
\end{array} \right)\otimes U_{ie} + 
\left( \begin{array}{rr} e^{i\chi} & 0 \\ 0 & 0  
\end{array} \right)\otimes 1_{i}\otimes  
F_{0_e \rightarrow \mu_e} , 
\end{eqnarray} 
where the first matrix in each term represents the spatial part and
the operator $F_{0_e \rightarrow \mu_e}$ flips
$|0_e \rangle$ to $|\mu_e \neq 0_e \rangle$. Fig.~\ref{fig:CPNetwork} 
shows the equivalent quantum network. The interference pattern is 
determined by
\begin{eqnarray} 
\nu_{\mu} e^{i\alpha_{\mu}} & = &  
{\text{Tr}}_{ie}\big[U_{ie} \rho|0_{e}  
\rangle \langle 0_{e}|  
F_{0_e \rightarrow \mu_e}^{\dagger} \big]  
\nonumber \\ 
 & = & {\text{Tr}}_{ie}\big[ U_{ie} \rho 
|0_{e} \rangle \langle \mu_{e}| \big] =  
{\text{Tr}}_{i} \big[ m_{\mu} \rho \big]  
\label{eq:cpint2} 
\end{eqnarray}  
for each $\mu = 1, \ldots, K-1$. The set $\{ \nu_{\mu}
e^{i\alpha_{\mu}} \}$, $\mu = 0, \ldots , K-1$, contains maximal 
information about the interference
effect associated with the CP map, by measuring on the system alone. For
unitarily evolving mixed states one obtains $\nu_{\mu} = 
\delta_{0 \mu}$, due to orthogonality of the environmental states, and the 
surviving interference pattern Eq.~(\ref{eq:cpint}) reduces to that of 
\cite{sjoqvist00}.

The results above can be derived by considering purifications. 
We may lift $\varrho$ in Eq.~(\ref{eq:rho}) to a purified 
state $|\Psi \rangle$ by attaching an ancilla according to
\begin{equation}
|\Psi \rangle=
\sum_k \sqrt{w_{k}} |k_i \rangle |0_e \rangle |k_a \rangle 
\end{equation} 
with $\{ |k_a \rangle \}$ a basis in an auxiliary Hilbert space 
of dimension at least as large as that of the internal Hilbert 
space. This state is mapped by the operators $U = U_{ie}  
\otimes I_a$ and $F = I_{i} \otimes F_{0_e \rightarrow \mu_e} 
\otimes I_a$ in the target and reference beam, respectively, i.e.  
\begin{eqnarray}
|\Psi_{\text{tar}} \rangle & = & 
\sum_k \sqrt{w_{k}} \big[ U_{ie} |k_i \rangle 
|0_e \rangle \big] |k_a \rangle , 
\nonumber \\ 
|\Psi_{\text{ref}} \rangle & = & 
\sum_k \sqrt{w_{k}} |k_i \rangle \big[  F_{0_e \rightarrow \mu_e} 
|0_e \rangle \big] |k_a \rangle , 
\end{eqnarray}
Their inner-product becomes  
\begin{eqnarray}
\langle \Psi_{\text{ref}} |\Psi_{\text{tar}} \rangle = 
\sum_k w_{k} \langle k_i | \langle 0_e | 
F_{0_e \rightarrow \mu_e}^{\dagger} U_{ie} |0_e \rangle | k_i \rangle   
\nonumber \\
= \sum_k w_{k} \langle k_i | m_{\mu} |k_i \rangle =
{\text{Tr}}_{i} ( m_{\mu} \rho ) 
\label{eq:purification}
\end{eqnarray}
in agreement with Eqs. (\ref{eq:cpint}) and (\ref{eq:cpint2}). 
 
To illustrate the above, let us consider the depolarization channel
\cite{preskill} acting on a qubit in the initial state
$\rho=\frac{1}{2}(I+\vec{r}\cdot\vec{\sigma})$, where
$\vec{r}=(x,y,z)$ is the Bloch vector with the length $|\vec{r}|\leq 1$,
$\vec{\sigma} = (\sigma_{x},\sigma_{y},\sigma_{z})$ are the
standard Pauli matrices, and $I$ is the $2\times 2$ unit matrix. We will model
this with the Kraus operators
\begin{eqnarray} 
m_{0} = \sqrt{1-p}~I , & & 
m_{1} = \sqrt{p/3}~\sigma_{x} , \nonumber \\ 
m_{2} = \sqrt{p/3}~\sigma_{y} , & & 
m_{3} = \sqrt{p/3}~\sigma_{z}   
\label{eq:depolkraus}
\end{eqnarray} 
that map $\rho \rightarrow \rho' = \frac{1}{2} (I+\vec{r}'  
\cdot\vec{\sigma})$. Here, $m_{1}$, $m_{2}$,  
and $m_{3}$, correspond to bit flip, both bit and phase flip,  
and phase flip, respectively. $p$ is the probability that  
one of these errors occurs and it determines the shrinking factor  
$|\vec{r}'|/|\vec{r}|=(1-4p/3)$ of the Bloch vector. Exposing  
the depolarization channel to one of the interferometer beams,  
the interference pattern is determined by Eq. (\ref{eq:cpint})  
as 
\begin{eqnarray} 
\nu_{0} e^{i\alpha_{0}} =  
{\text{Tr}}_{i}(\rho m_{0}) = \sqrt{1-p}. 
\end{eqnarray} 
This quantity is real and positive so that $\alpha_{0}=0$,  
thus the channel only reduces the visibility by the factor  
$\sqrt{1-p}$. If the state of the environment is flipped to any of the  
other $|\mu_{e} \rangle$'s the interference pattern  
Eq.~(\ref{eq:cpint2}) is determined by  
\begin{eqnarray} 
\nu_{\mu} e^{i\alpha_{\mu}} =  
{\text{Tr}}_{i}(\rho m_{\mu}) = \sqrt{\frac{p}{3}}~r_{\mu} ,  
\end{eqnarray}  
where $r_{\mu}=x,y,z$ for $\mu =1,2,3$, respectively. These are  
also real and positive so that $\alpha_{\mu} = 0$. The only  
effect is a nonvanishing visibility proportional to the   
probability amplitude $\sqrt{p/3}$ associated with the  
corresponding error. The absence of phase shifts can be  
understood from the fact that the depolarization channel  
only shrinks the length of the Bloch vector.  
 
Another illustration of the general formalism above is provided  
by the amplitude damping channel \cite{preskill}, which models,  
e.g., the decay of an atom from one of its excited states to  
its ground state by emitting a photon. It may be described by the
Kraus operators
\begin{eqnarray} 
m_{0} & = & \frac{1}{2} (I + \sigma_{z}) + 
\frac{\sqrt{1-p}}{2} (I - \sigma_{z}) , \nonumber \\ 
m_{1} & = & \frac{\sqrt{p}}{2} 
(\sigma_{x}+i\sigma_{y}), 
\end{eqnarray} 
where $p$ is the decay probability. When no photon has been  
emitted in the reference beam the interference pattern is  
determined by  
\begin{equation} 
\nu_{0} e^{i\alpha_{0}} =  
{\text{Tr}}_{i}(\rho m_{0})=\frac{1}{2} 
\left(1+\sqrt{1-p}+z[1-\sqrt{1-p}]\right),  
\label{eq:ampdamp1}
\end{equation} 
which is real and positive so that $\alpha_{0}=0$, and only 
the visibility $\nu_{0}$ is affected anisotropically by the  
channel. On the other hand, if a photon has been emitted in the  
reference beam, we obtain
\begin{equation}
\nu_{1} e^{i\alpha_{1}} = {\text{Tr}}_{\text {i}}(\rho m_{1}) =  
\frac{\sqrt{p}}{2} (x+iy) ,  
\label{eq:ampdamp2}
\end{equation}
being dependent only upon the $x-y$ projection of the initial  
Bloch vector. Here, the visibility decrease is proportional to  
the decay probability amplitude $\sqrt{p}$ and there is a shift  
in the interference oscillations determined by the angle of the  
$x-y$ projection to the $x$ axis. By checking whether or not 
there is a photon in the reference beam it should be possible 
experimentally to distinguish the two interference patterns determined 
by Eqs.~(\ref{eq:ampdamp1}) and (\ref{eq:ampdamp2}) by post-selection. 
 
In the general case, it is convenient to make a polar decomposition 
of $m_{\mu}$ such that  
\begin{equation} 
m_{\mu} = h_{\mu} u_{\mu} , 
\label{eq:m0} 
\end{equation} 
where $h_{\mu}$ is Hermitian and positive, and $u_{\mu}$ is unitary.  
The action of each $m_{\mu}$ is uniquely defined up to $N$ phase  
factors by the evolution of the system's density operator.  
This ambiguity must be associated with the  
corresponding unitarity $u_{\mu}$ as the Hermitian part is 
unique.  
 
Let us clarify this point for a qubit, where we may write $u_{\mu} =
e^{-i\theta_{\mu} \vec{e}_{\mu} \cdot\vec{\sigma}}$ and $h_{\mu} = a_{\mu} +
\vec{b}_{\mu} \cdot \vec{\sigma}$. It is possible to evaluate
Eq.~(\ref{eq:cpint2}) for $\rho=\frac{1}{2}(I+\vec{r} \cdot \vec{\sigma})$
yielding
\begin{eqnarray} 
\nu_{\mu} e^{i\alpha_{\mu}} & = & (a_{\mu} + \vec{r} \cdot  
\vec{b}_{\mu}) \cos \theta_{\mu} + (\vec{r}\times\vec{e}_{\mu})  
\cdot \vec{b}_{\mu} \sin\theta_{\mu}   
\nonumber \\  
 & & -i (\vec{e}_{\mu} \cdot \vec{b}_{\mu} + a_{\mu}\vec{r} \cdot  
\vec{e}_{\mu}) \sin \theta_{\mu} .  
\end{eqnarray}  
Now, with $\vec{r} = r\vec{n}$ the density operator is unaffected by the
change $u_{\mu} \rightarrow u_{\mu} e^{-i\gamma_{\mu} \vec{n} \cdot
  \vec{\sigma}}$ as the additional part commutes with $\rho$. However, the
interference pattern is determined by this new unitarity and therefore changes
according to
\begin{eqnarray}  
\cos \theta_{\mu} & \rightarrow &  
\cos \tilde{\theta}_{\mu} = \cos \theta_{\mu} \cos \gamma_{\mu} -  
\vec{e}_{\mu} \cdot \vec{n} \sin \theta_{\mu} \sin \gamma_{\mu}  
\nonumber \\  
\vec{e}_{\mu} \sin \theta_{\mu} & \rightarrow &  
\tilde{\vec{e}}_{\mu} \sin \tilde{\theta}_{\mu} =  
\vec{n} \cos \theta_{\mu} \sin \gamma_{\mu}  
\nonumber \\  
 & & + \vec{e}_{\mu}  
\sin \theta_{\mu} \cos \gamma_{\mu} + \vec{e}_{\mu} \times  
\vec{n} \sin \theta_{\mu} \sin \gamma_{\mu} ,  
\end{eqnarray} 
where $u_{\mu} e^{-i\gamma_{\mu} \vec{n} \cdot \vec{\sigma}} =
e^{-i\tilde{\theta}_{\mu} \tilde{\vec{e}}_{\mu} \cdot \vec{\sigma}}$.
 
The phase ambiguity can be removed by introducing a notion of  
parallel transport and concomitant geometric phases for a 
continuous (time) parametrization of the CP map. For each 
interference pattern, the $N$ parallel transport conditions 
read  
\begin{equation} 
\langle k| \tilde{u}_{\mu}^{\dagger} (t) \dot{\tilde{u}}_{\mu} (t) 
|k\rangle=0,  
\hskip1cm  
k=1, \ldots N ,  
\label{eq:ptc}  
\end{equation} 
where we have decomposed $m_{\mu} (t) = h_{\mu} (t) u_{\mu} (t) = 
h_{\mu} (t) v_{\mu} \tilde{u}_{\mu} (t)$ with $v_{\mu} =u_{\mu} (0)$ 
completely specified by the channel, $\tilde{u}_{\mu}(0)=1$, and we 
have assumed $h_{\mu} (0) = \delta_{\mu 0}$. These conditions naturally 
extend those in Eq.~(13) of \cite{sjoqvist00} to the case of CP maps. 
They are sufficient and necessary to arrive at a unique notion of 
geometric phase in the context of single-particle interferometry. 
The set of these geometric phases for $\mu=0,\ldots,K-1$ provides 
the complete geometric picture of the CP map in interferometry,
given the unitary representation. 
 
Now, let us assume for simplicity that $h_{\mu}$ and $\rho$  
diagonalize in the same basis $\{ |k\rangle \}$ for all  
$t\geq 0$. In the cyclic case, where for each $k$ we have  
$\tilde{u}_{\mu}: |k\rangle \rightarrow e^{i\beta_{k}^{\mu}} |k\rangle$,  
$\beta_{k}^{\mu}$ being the corresponding cyclic pure state  
geometric phase, the interference pattern is given by 
\begin{eqnarray} 
\nu_{\mu}e^{i\Phi_{\mu}} = \sum_{k} w_{k}  
\langle k|h_{\mu}|k\rangle \langle k|v_{\mu}|k\rangle  
e^{i\beta_{k}^{(\mu)}} , 
\label{eq:gpcyclic}
\end{eqnarray} 
where all $\langle k|h_{\mu}|k\rangle \geq 0$ are real-valued  
as $h_{\mu}$ is Hermitian and positive. 
 
Let us use Eq.~(\ref{eq:gpcyclic}) to compute the geometric phase
for a qubit with $r\neq 0$ in the depolarization channel with a
unitary rotation added. Thus, we replace the Kraus operators $m_{\mu}$
in Eq.~(\ref{eq:depolkraus}) by $m_{\mu} \tilde{u}$ with $\tilde{u}$
being an $SU(2)$ operator that fulfills the parallel transport
conditions Eq.~(\ref{eq:ptc}).  Here, $h_{\mu}$ is diagonal for all
Kraus operators, $(v_{1},v_{2},v_{3}) =
(\sigma_{x},\sigma_{y},\sigma_{z})$, and $\tilde{u}_{\mu} = \tilde{u}$
for all $\mu$. For cyclic $\tilde{u}$, the interference patterns are
determined by
\begin{eqnarray} 
\nu_{0}e^{i\Phi_{0}}&=&\sqrt{1-p}(\cos(\Omega/2)+ir\sin(\Omega/2)) 
\nonumber \\ 
\nu_{1}e^{i\Phi_{1}}&=& 0\nonumber \\ 
\nu_{2}e^{i\Phi_{2}}&=& 0\nonumber \\ 
\nu_{3}e^{i\Phi_{3}}&=&\sqrt{p/3}(r\cos(\Omega/2)+i\sin(\Omega/2))    
\end{eqnarray} 
with $\Omega$ the solid angle enclosed by the loop on the  
Bloch sphere. The first interference pattern is precisely that  
obtained in \cite{sjoqvist00} modified by a visibility factor  
$\sqrt{1-p}$. $\nu_{1}$ and $\nu_{2}$ vanish since the  
corresponding errors involve bit flips. Surprisingly, the pure  
phase flip in the last interference pattern introduces a  
nontrivial change in the appearance of $r$. This is due to  
the fact that this phase flip introduces an additional relative 
sign between the weights of the pure state interference patterns 
and is purely an effect of the decoherence.  
 
We may also consider the case where $\tilde{u}$ takes 
$|0\rangle \rightarrow |1\rangle$ and vice versa.  
Any such $\tilde{u}$ fulfilling the parallel transport conditions  
Eq. (\ref{eq:ptc}) may be written as $e^{-i(\pi /2) [\cos \varphi  
\sigma_{x} + \sin \varphi \sigma_{y}]}$.  
Here, only the errors containing bit flips produce a  
nonvanishing interference effect that can be interpreted  
geometrically by noting that the $SU(2)$ error operators $iv_{1}$  
and $iv_{2}$ are themselves taking $|0 \rangle \rightarrow |1 \rangle$ and  
$|1 \rangle \rightarrow |0 \rangle$ along geodesics intersecting the $y$  
and $x$ axis, respectively. Thus, $iv_{1} \tilde{u}$ defines a  
closed loop on the Bloch sphere that encloses the solid  
angle $2\pi - 2\varphi$, yielding  
\begin{eqnarray} 
\nu_{1}e^{i\Phi_{1}} & = &  
-i \sqrt{\frac{p}{3}} \Big[ \frac{1+r}{2}  
\langle 0|iv_{1} \tilde{u}| 0 \rangle + \frac{1-r}{2}   
\langle 1|iv_{1} \tilde{u}| 1 \rangle \Big]  
\nonumber \\    
 & = & \sqrt{\frac{p}{3}} e^{-i\pi /2} (\cos \varphi + ir\sin \varphi) .  
\end{eqnarray} 
Similarly, $iv_{2} \tilde{u}$ defines the solid angle   
$3\pi - 2\varphi$, so that the interference pattern is  
determined by     
\begin{eqnarray} 
\nu_{2}e^{i\Phi_{2}} & = &  
-i \sqrt{\frac{p}{3}} \Big[ \frac{1+r}{2}  
\langle 0|iv_{2} \tilde{u}| 0 \rangle + \frac{1-r}{2}   
\langle 1|iv_{2} \tilde{u}| 1 \rangle \Big]  
\nonumber \\    
& = &  \sqrt{\frac{p}{3}} e^{-i\pi} ( r\cos\varphi + i\sin\varphi) .  
\end{eqnarray} 
Again, there is a nontrivial change in the appearance of $r$  
in the last expression due to the fact that this error also  
contains a phase flip. 

To summarize, we have provided a generalization of the notion  
of relative phase to completely positive maps with known 
unitary representation, based on interferometry. We have 
further introduced parallel transport conditions that define 
the geometric phase for such maps. The interference effect is 
embodied in a set of interference patterns defined by flipping 
the environment state in one of the two particle beams. We have 
shown in the qubit case that this structure gives rise to 
interesting additional information about the geometry of the 
evolution defined by  the CP map. We hope that this work will 
trigger new experiments on geometric phases for quantal systems 
exposed to environmental interactions.

We would like to thank Artur Ekert for useful discussions.  The work by E.S.
was financed by the Swedish Research Council. D.K.L.O acknowledges the support
of CESG (UK) and QAIP grant IST-1999-11234.
 
{\it Note added:} After completing this work, it has come to  
our knowledge that Peixoto de Faria {\it et al.} \cite{peixoto02}  
have arrived at the interference pattern described  by Eq.  
(\ref{eq:cpint}) in the context of measurement theory.

\end{document}